\documentstyle[preprint,aps]{revtex}
\input psfig.tex

\begin{document}

\title{Pion correlation 
from Skyrmion-AntiSkyrmion annihilation} 

\author{Yang Lu and R.~D.~Amado}
\address{Department of Physics,  \\
University of Pennsylvania, Philadelphia, PA 19104, USA  \\}

\date{July 13, 1995}

\maketitle

\begin{abstract}
We study two pion correlations from 
Skyrmion and antiSkyrmion collision, using the product 
ansatz and an approximate random grooming method 
for nucleon projection. The spatial-isospin coupling inherent
in the Skyrme model,  along with empirical averages, leads to
correlations not only among pions of like charges but 
also among unlike charge types.
\end{abstract}


\section{Introduction}

We have studied nucleon--antinucleon annihilation at rest in 
the large $N_c$ limit of QCD in a series of papers \cite{ACDLS,SA,LA}. The 
gross features of annihilation  emerge correctly from this 
picture. These features include the pion momentum distribution, number and 
charge spectra and branching ratios into meson channels.
In these studies the initial baryon density distribution was
taken as spherically symmetric and only the subsequent
evolution of the mesons was treated dynamically.  In the
large $N_c$ limit this dynamics is the classical
dynamics of Skyrme like models.  To discuss annihilation in
flight, pion correlations, the effects of the spin and isospin
of the individual baryons and the initial state dynamics
involves considerable improvement and sophistication of
our previous starting point. 
We also need to
address the question of quantum, or finite $N_C$, 
corrections. These are usually introduced by rotating the Hedgehog 
(and antiHedgehog) slowly and quantizing the rotational 
degrees of freedom.  In the Skyrme picture of 
$\overline{N}N$ collision, we need to collide a rotating
Skyrmion with a rotating antiSkyrmion, 
project out the nucleon and antinucleon states, and
follow the dynamical evolution of the classical system to 
pion radiation.
This is, of course, a much more demanding task than the 
study of classical dynamics of annihilation
without grooming \cite{Seki}, which is already an
elaborate and numerically challenging calculation.
These difficult issues of classical Skyrmion dynamics and
quantum correction in annihilation bear 
strongly on pion Bose-Einstein correlations\cite{ACDLL}, charge 
asymmetry\cite{Hasan,Myhrer3} in $\overline{p}p\rightarrow\pi^+\pi^-$ 
and annihilation in flight in general.

This paper is a first modest step in the study of annihilation in flight and
of pion intensity  correlations.  
We study
annihilation in flight with the product
ansatz and with no subsequent pion dynamics.  
Pion intensity correlations emerge as a consequence of random 
grooming of approaching Skyrmion and antiSkyrmion. 
The result are similar to the commonly considered
Bose-Einstein effects resulting from thermal  decoherence. However, 
the correlation in the present study results from the spatial-isospin
coupling inherent in Skyrme soliton.  Pion correlations from this
spatial-isospin coupling  were studied by Blaizot and Diakonov
\cite{BD} in heavy ion collision. 

In the next section, we obtain slowly moving Skyrmion configurations
from Galilean transformation. We approximate the full dynamics by
the configuration of a Skyrmion and antiSkyrmion nearing each 
other. Such a configuration at the point of closest
approach is used in Section
3 to obtain a coherent state amplitude describing the pion
radiation from annihilation. Interaction between the
Skyrmion and antiSkyrmion or among the pions is not considered.
In section 4, we discuss the random grooming of the Skyrmion 
and antiSkyrmion sources and resulting pion correlations. Section 5
outlines the case of Lorentz transformed Skyrmions and compares 
the results with those from the Galilean case. Section 6 
gives our results and prospect for future development. The various
approximations we make, some of them quite drastic, are discussed
in the sections where they are employed.  

\section{Classical moving Hedgehogs: Galilean case}

The large $N_c$ dynamics we consider is the Skyrme model \cite{Skyrme}
which is a classical nonlinear 
field theory of pions only in which the 
baryons emerge as nonperturbative topological solitons.  These baryons
are not nucleons but rather hedgehogs (so called because the pion isospin
and spatial directions are correlated).  The hedgehogs are a superposition
of all possible baryonic states with isospin equal to spin. The nucleon,
or antinucleon, is then projected out from the hedgehog configuration
by grooming or rotating the hedgehog.
When a hedgehog and antihedgehog approach each other, there is a complicated
interaction coming from the nonlinear nature of the Skyrme dynamics,
leading finally to pion radiation.  This at least is what happens for
close collisions, and has been seen in the one full calculation that 
has been done. \cite{Seki} The nonlinear nature of the Skyrme lagrangian
and the presence of fourth order derivatives makes the calculation
numerically unstable.  Thus a full investigation of the nature of
Skyrmion antiSkyrmion collisions as a function of energy and impact
parameter and of the subsequent pion radiation has not yet been
carried out even for the hedgehog anti-hedgehog orientation, let alone
for groomed hedgehogs. We do not even try.  Rather we take as the
initial state a hedgehog and antihedgehog moving in opposite directions and
at their point of closest approach.  This is the product ansatz for
moving hedgehogs. The hedgehogs themselves we take as solutions of the
Skyrme equations.  We then use their classical 
pion configuration as the source of
pions for our quantum coherent state with no further dynamics among those
pions. The calculation of the Cal Tech group \cite{Seki} and the
one model calculation we carried out in simple geometry \cite{ASW}
suggest that once annihilation begins, it proceeds very rapidly, at
the causal limit, and that the subsequent pion wave in the radiation zone
looks very much like the wave one would obtain from a free wave fourier
transform of the source.  We use these facts and the fact that anything
else is too difficult for now to justify our simple approximations.

In the center of mass, we represent the colliding 
nucleon and antinucleon by
a classical Skyrmion and antiSkyrmion approaching each other 
with velocity $\mbox{\boldmath $v$}$ and impact parameter 
$\mbox{\boldmath $b$}$.\footnote{ 
Bold face letters denote 3-vectors. Overhead arrows denote 
iso-vectors.}
We make the approximation that annihilation and the subsequent
pion radiation comes 
at the instance of closest approach, as shown in Fig. 1. 
Final state interactions
among the resulting pions are neglected as discussed above.  

 The pion field configuration of a  moving hedgehog is 
the Lorentz boosted static hedgehog. For small velocities,
we can approximate it by a Galilei boosted hedgehog, which
we study first.

 The classical field of pions in Fig.~1 can be written as 
\begin{eqnarray}
\vec{\Phi}(\mbox{\boldmath $r$},t)&= &
\vec{\Phi}_a (\mbox{\boldmath $r$}, t) + 
\vec{\Phi}_b (\mbox{\boldmath $r$}, t)  
\nonumber \\
\vec{\Phi}_a (\mbox{\boldmath $r$}, t) &=& 
 (\mbox{\boldmath $r$}- \mbox{\boldmath $v$} t+\mbox{\boldmath $b$}/2) 
 g(| \mbox{\boldmath $r$}- \mbox{\boldmath $v$} t+\mbox{\boldmath $b$}/2|)  
 \nonumber \\
\vec{\Phi}_b (\mbox{\boldmath $r$}, t) &=&
- (\mbox{\boldmath $r$} + \mbox{\boldmath $v$} t-\mbox{\boldmath $b$}/2) g(|
\mbox{\boldmath $r$} + \mbox{\boldmath $v$} t-\mbox{\boldmath $b$}/2|) 
\label{eq:SantiS}
\end{eqnarray}
where $g(r)=f(r)/r$ ($f(r)$ is the usual Skyrme profile function, modulo
$f_\pi$), $a$ is the Skyrmion and $b$ is 
the antiSkyrmion. The fact that the iso-vector fields on the left 
in (\ref{eq:SantiS}) are written in terms of radial spatial vectors 
on the right is the hedgehog feature.

\section{Requantizing}

To form a coherent state, we need to decompose 
the classical pion field (\ref{eq:SantiS}) into a plane wave and 
pick out the positive frequency part.
We neglect the pion interactions and thus  
the pion wave satisfies the free wave 
equation (the isospin index is suppressed for now), 
\begin{equation}
\nabla^2 \psi(\mbox{\boldmath $r$},t) 
=( \frac{\partial^2}{\partial t^2 }+m^2)
\psi(\mbox{\boldmath $r$},t)
\end{equation}
where $m$ is the pion mass.
  
We introduce the  momentum space amplitude 
$h(\mbox{\boldmath $k$}, \omega)$ by
\begin{equation}
\psi(\mbox{\boldmath $r$},t)= \int d^3k d\omega
 e^{i\mbox{\boldmath $k$}\cdot \mbox{\boldmath $r$}} e^{-i
\omega t} h(\mbox{\boldmath $k$}, \omega)
\end{equation}
and we find
\begin{equation}
h(\mbox{\boldmath $k$}, \omega) =
 \delta(k^2 +m^2-\omega^2)G(\mbox{\boldmath $k$}, \omega)
\end{equation}
and therefore 
\begin{equation}
\psi(\mbox{\boldmath $r$},t)= \int d^3k
 d\omega( e^{ i\mbox{\boldmath $k$ }\cdot \mbox{\boldmath $ r$}} e^{-i
\omega_k t} G(\mbox{\boldmath $k$}, \omega_k)/(2 \omega_k) 
+ e^{-i\mbox{\boldmath $k$}\cdot \mbox{\boldmath $ r$}} e^{i
\omega_k t} G(\mbox{\boldmath $k$},- \omega_k)/(2 \omega_k)).
\end{equation}
The positive frequency part of the wave comes from the first term,
the one with $e^{-i \omega_k t}$ and
therefore we need $G(\mbox{\boldmath $k$}, \omega_k)$,  
($\omega_k = \sqrt{k^2+m^2}$).
From the field and its rate of change at $t=0$, the instant of
nearest approach, we obtain
\begin{eqnarray}
\psi(\mbox{\boldmath $r$}, 0) & = & 
\int d^3k e^{i \mbox{\boldmath $k$}\cdot \mbox{\boldmath $r$}}
\frac{G(\mbox{\boldmath $k$}, \omega_k) + G(\mbox{\boldmath $k$}, 
-\omega_k)}{2 \omega_k}
\nonumber \\
  & = & \int d^3k  e^{i \mbox{\boldmath $k$}\cdot \mbox{\boldmath $r$}} 
\gamma_1(\mbox{\boldmath $ k$})
  \label{eq:gamma1}
\end{eqnarray}
and
\begin{eqnarray}
\frac {d \psi(\mbox{\boldmath $r$}, 0)}{dt}  & = & 
\int d^3k e^{i \mbox{\boldmath $k$}\cdot \mbox{\boldmath $r$}}
(-i)\frac{G(\mbox{\boldmath $k$}, \omega_k) - G(\mbox{\boldmath $k$}, 
-\omega_k)}{2 }
 \nonumber  \\
& = & \int d^3k  
e^{i \mbox{\boldmath $k$}\cdot \mbox{\boldmath $r$}} 
\gamma_2(\mbox{\boldmath $k$})
  \label{eq:gamma2}
\end{eqnarray}
We therefore see that
\begin{equation}
G(\mbox{\boldmath $k$}, \omega_k) = \omega_k \gamma_1(\mbox{\boldmath $k$}) 
+i \gamma_2(\mbox{\boldmath $k$})
\label{eq:gk}
\end{equation}
and the fourier amplitude of the field that we need is given by
\begin{equation}
h(\mbox{\boldmath$ k$}, \omega_k) = \frac{G(\mbox{\boldmath $k$}, 
\omega_k)}{2 \omega_k}
\label{eq:h}
\end{equation}

To obtain $\gamma_1$, $\gamma_2$ and thus $G(k)$, 
we fourier transform the field configuration (\ref{eq:SantiS}) 
and its time derivative 
\begin{equation}
\vec{\gamma}_1(\mbox{\boldmath $k$})
=\frac{1}{(2\pi)^3}
\int d^3r e^{-i \mbox{\boldmath $k$}\cdot\mbox{\boldmath $r$}}\vec{\Phi} 
(\mbox{\boldmath $r$},0)
\label{eq:Fourier1}
\end{equation}
\begin{equation}
\vec{\gamma}_2(\mbox{\boldmath $k$})
=\frac{1}{(2\pi)^3}
\int d^3r e^{-i \mbox{\boldmath $k$}\cdot\mbox{\boldmath $r$}}
\frac{d\vec{\Phi}}{dt} (\mbox{\boldmath $r$},0)
\label{eq:Fourier2}
\end{equation}
We split the amplitude into
its contributions from the hedgehog and from the antihedgehog.
\begin{equation}
\gamma\rightarrow\gamma_a+\gamma_b
\end{equation}
\begin{eqnarray}
\vec{\gamma}_{1a}&=&\frac{1}{(2\pi)^3}
\int d^3r e^{-i \mbox{\boldmath $k$}\cdot\mbox{\boldmath $r$}}
(\mbox{\boldmath $r$}+\frac{\mbox{\boldmath $b$}}{2})
g(|\mbox{\boldmath $r$}
+\frac{\mbox{\boldmath $b$}}{2}|)
\\
\vec{\gamma}_{1b}&=&-\frac{1}{(2\pi)^3}
\int d^3r e^{-i \mbox{\boldmath $k$}\cdot\mbox{\boldmath $r$}}
(\mbox{\boldmath $r$}-\frac{\mbox{\boldmath $b$}}{2})
g(|\mbox{\boldmath $r$}-\frac{\mbox{\boldmath $b$}}{2}|)
\end{eqnarray}
Introducing 
\begin{equation}
G(k^2)=\frac{1}{(2\pi)^3}\int d^3\rho 
e^{-i \mbox{\boldmath $k$}\cdot\mbox{\boldmath $\rho$}}
 g(|\mbox{\boldmath $\rho$}|)
\end{equation}
it is easy to show that 
\begin{eqnarray}
\vec{\gamma}_{1a}&=& 2i\mbox{\boldmath $k$} 
e^{i\mbox{\boldmath $k$}\cdot{\frac{\mbox{\boldmath $b$}}{2}}} G'(k^2)
\\
\vec{\gamma}_{1b}&=& -2i\mbox{\boldmath $k$} 
e^{-i\mbox{\boldmath $k$}\cdot{\frac{\mbox{\boldmath $b$}}{2}}} G'(k^2)
\end{eqnarray}
with $G'(k^2)=dG/d(k^2)$.

\begin{eqnarray}
\dot{\vec{\Phi}}_a(\mbox{\boldmath $r$},0)&=&-\mbox{\boldmath $v$} 
g(|\mbox{\boldmath $r$}+\frac{\mbox{\boldmath $b$}}{2}|^2)
-(\mbox{\boldmath $r$}+\frac{\mbox{\boldmath $b$}}{2})\,
\mbox{\boldmath $v$} 
\cdot
2(\mbox{\boldmath $r$}+\frac{\mbox{\boldmath $b$}}{2})\,
g'(|\mbox{\boldmath $r$}+\frac{\mbox{\boldmath $b$}}{2}|^2)
\\
\dot{\vec{\Phi}}_b(\mbox{\boldmath $r$},0)&=&-\mbox{\boldmath $v$} 
g(|\mbox{\boldmath $r$}-\frac{\mbox{\boldmath $b$}}{2}|^2)
-(\mbox{\boldmath $r$}-\frac{\mbox{\boldmath $b$}}{2})\,
\mbox{\boldmath $v$} 
\cdot
2(\mbox{\boldmath $r$}-\frac{\mbox{\boldmath $b$}}{2})\,
\,
g'(|\mbox{\boldmath $r$}-\frac{\mbox{\boldmath $b$}}{2}|^2)
\end{eqnarray}
where 
\begin{equation}
g'(\mbox{\boldmath $\rho$}^2)=\frac{dg}{d\mbox{\boldmath $\rho$}^2}
\end{equation}

It is straightforward to show 
\begin{eqnarray}
\vec{\gamma}_{2a}&=&\frac{1}{(2\pi)^3}
\int d^3r e^{-i\mbox{\boldmath $k$}\cdot\mbox{\boldmath $r$}} 
\dot{\vec{\Phi}}_a(\mbox{\boldmath $r$},0)
\nonumber \\
&=& -\mbox{\boldmath $v$}e^{i\mbox{\boldmath $k$}
\cdot{\frac{\mbox{\boldmath $b$}}{2}}}G(k^2)
+\vec{\gamma}_{2a}^{II}
\end{eqnarray}
where
\begin{eqnarray}
\vec{\gamma}_{2a}^{II}&=&
\frac{1}{(2\pi)^3}
\int d^3r 
e^{-i\mbox{\boldmath $k$}\cdot\mbox{\boldmath $r$}}
(\mbox{\boldmath $r$}+\frac{\mbox{\boldmath $b$}}{2})
[ -2\mbox{\boldmath $v$}\cdot(\mbox{\boldmath $r$}
+\frac{\mbox{\boldmath $b$}}{2})]
g'(|\mbox{\boldmath $r$}+\frac{\mbox{\boldmath $b$}}{2}|)
\nonumber \\
&=& 
\frac{1}{(2\pi)^3} e^{i\mbox{\boldmath $k$}
\cdot{\frac{\mbox{\boldmath $b$}}{2}}}
\int d^3\rho e^{-i\mbox{\boldmath $k$}
\cdot{\mbox{\boldmath $\rho$}}}
\mbox{\boldmath $\rho$}(-2\mbox{\boldmath $v$}
\cdot\mbox{\boldmath $\rho$})g'(\rho^2)
\nonumber \\
&=&
-\frac{1}{(2\pi)^3} e^{i\mbox{\boldmath $k$}
\cdot{\frac{\mbox{\boldmath $b$}}{2}}}
\int d^3\rho e^{-i\mbox{\boldmath $k$}
\cdot{\mbox{\boldmath $\rho$}}}\mbox{\boldmath $v$}\cdot
\mbox{\boldmath $\rho$}
\nabla_{\bf{\rho}}g(\rho^2)
\nonumber \\
&=&\frac{1}{(2\pi)^3} e^{i\mbox{\boldmath $k$}
\cdot{\frac{\mbox{\boldmath $b$}}{2}}}
\int d^3\rho g(\rho^2)\nabla_{\bf{\rho}}[
e^{-i\mbox{\boldmath $k$}\cdot{\mbox{\boldmath $\rho$}}}
\mbox{\boldmath $v$}\cdot\mbox{\boldmath $\rho$}]
\nonumber \\
&=& e^{i\mbox{\boldmath $k$}\cdot{\frac{\mbox{\boldmath $b$}}{2}}} 
\mbox{\boldmath $v$} G(k^2)
+2e^{i\mbox{\boldmath $k$}\cdot{\frac{\mbox{\boldmath $b$}}{2}}} 
\mbox{\boldmath $k$}\; \mbox{\boldmath $v$}\cdot
\mbox{\boldmath $k$} G'(k^2)
\end{eqnarray}
Note we use 
$\nabla_{\mbox{\boldmath $\rho$}}g(\rho^2)=2\mbox{\boldmath $\rho$}
\frac{dg}{d\rho^2}$
and then integrate by parts. Thus 
the fourier amplitude (\ref{eq:h}) of the field of the Skyrmion we 
need
is 
\begin{equation}
\vec{h}_a(\mbox{\boldmath $k$},\omega_k)= i
\mbox{\boldmath $k$} e^{i\mbox{\boldmath $k$}
\cdot{\frac{\mbox{\boldmath $b$}}{2}}}G'(k^2)
(1+\frac{ \mbox{\boldmath $v$}\cdot\mbox{\boldmath $k$ }}{\omega_k})
\end{equation}
Similarly we have for the field from the antiSkyrmion
\begin{equation}
\vec{h}_b(\mbox{\boldmath $k$},\omega_k)= -i
\mbox{\boldmath $k$} e^{-i\mbox{\boldmath $k$}
\cdot{\frac{\mbox{\boldmath $b$}}{2}}}G'(k^2)
(1-\frac{ \mbox{\boldmath $v$}\cdot\mbox{\boldmath $k$}}{\omega_k})
\end{equation}
A quantum coherent state represent the configuration in Fig.~1
can then be written as 
\begin{equation}
|h\rangle =\exp [\int d^3k (\vec{h}_a(\mbox{\boldmath $k$},\omega_k) +
\vec{h}_b(\mbox{\boldmath $k$},\omega_k))
\cdot \vec{a}^+(\mbox{\boldmath $k$})]|0\rangle
\label{eq:coherentstate}
\end{equation}

\section{Random grooming and Correlations}
 
 We next consider grooming
of the hedgehog and anti-hedgehog to introduce
spin and isospin. In fact we will argue that
we do not have to quantize the grooming but that
we can consider random grooming.  
This is justified 
by consideration of time scales.  Spin and 
isospin are introduced by spinning or rotating the
hedgehog at a fixed rate. 
That rate of rotation
is slow compared to the annihilation time.  
Thus during any given annihilation the groomed
hedgehogs do not rotate much.  But if
there is no polarization, their orientation is
random at the moment of annihilation and it is 
that random orientation or random grooming that
we average over when we consider many events.

As shown in the calculation 
of Skyrmion and antiSkyrmion annihilation \cite{Seki,ASW},
the pions emerge promptly from the interaction region. 
This happens in about
the time for light signal to travel across the Skyrmion, 
which is $T_c=1/m $. Here $m $ is the pion mass.
The rotation period can be estimated from the $N-\Delta$ 
splitting which is of the order of $m$. The rotation energy
is $E_r\approx Ir^2\omega^2\approx M/(m^2 T_r^2)\sim m$, 
thus $T_r=1/m \sqrt{M/m}$. $M$ is the nucleon mass. The 
ratio of rotation to annihilation time is then $T_r/T_c\sim 
\sqrt{M/m}\approx 3$. If we were to estimate the rotation time by
the angular momentum (of order 1, noting $\hbar=1$), we would have
$J\sim
I\omega\sim M/m^2/T_r\sim 1$ and $T_r=M/m^2$. The ratio 
would be $T_r/T_c\sim M/m \approx 7$. 
Both these estimates suggest that the approximation of taking
a fixed orientation  during the fast annihilation is a 
reasonable one.  This is something that should be improved
in more detailed calculations.
In the limit of slow rotation, 
the isospin orientation of each Skyrmion can be considered as random 
between one annihilation event and another.
Experimental observables should then be the average over all
the different directions in isospin space.

To proceed with random grooming, we write the amplitudes 
$h(\mbox{\boldmath $k$})$
in spherical components
\begin{eqnarray} 
h_{a\mu}& = & \sqrt{\frac{4\pi}{3}}Y_{1\mu}(\hat{k})   
H_a(\mbox{\boldmath $k$}) 
\nonumber \\
h_{b\mu}& = & \sqrt{\frac{4\pi}{3}}Y_{1\mu}(\hat{k})   
H_b(\mbox{\boldmath $k$}) 
\end{eqnarray}
with
\begin{eqnarray}
H_a(\mbox{\boldmath $k$})& = & i
e^{i\mbox{\boldmath $k$}\cdot{\frac{\mbox{\boldmath $b$}}{2}}} k G'(k^2)
(1+\frac{\mbox{\boldmath $v$}\cdot\mbox{\boldmath $k$}}{\omega_k})
\nonumber \\
H_b(\mbox{\boldmath $k$})& = & -i
e^{-i\mbox{\boldmath $k$}\cdot{\frac{\mbox{\boldmath $b$}}{2}}} k G'(k^2)
(1-\frac{\mbox{\boldmath $v$}\cdot\mbox{\boldmath $k$}}{\omega_k})
\label{eq:Hab}
\end{eqnarray}
Independent grooming  of  the Skyrmion  and anti-Skyrmion 
means identifying  the total amplitude with the $\mu$-th component
of isospin as 
\begin{equation}
h_{\mu}(\mbox{\boldmath $k$},\Omega,\Xi)=\sqrt{\frac{4\pi}{3}} \left[
{\cal D}_{\mu\nu}^1 (\Omega) Y_{1\nu}(\hat{k}) 
H_a(\mbox{\boldmath $k$}) 
+ {\cal D}_{\mu\nu}^1 (\Xi) Y_{1\nu}(\hat{k}) 
H_b(\mbox{\boldmath $k$}) \right]
\label{eq:hgroomed}
\end{equation}
with $\Omega$ and $\Xi$ being two sets of independent Euler angles.
Observables calculated from the coherent state 
(\ref{eq:coherentstate}) with the amplitude (\ref{eq:hgroomed})
are to be averaged over the range of $\Omega$ and $\Xi$.

  First we examine the one-body rate, that is the probability for 
finding a pion of momentum $\mbox{\boldmath $k$}$ and isospin type $\mu$
in the coherent state. We have 
\begin{eqnarray}
W^{(1)}(\mu, \mbox{\boldmath $k$}) 
&=&\frac{1}{(8\pi^2)^2}\int d\Omega d\Xi
\; h_{\mu}^* (\mbox{\boldmath $k$},\Omega,\Xi)
\;h_{\mu} (\mbox{\boldmath $k$},\Omega,\Xi)
\nonumber \\
&=& \frac{1}{48\pi^3}\int d\Omega d\Xi
\nonumber \\
&& \left[
{\cal D}^{1*}_{\mu\nu}(\Omega)
{\cal D}^{1}_{\mu\nu'}(\Omega)
H^*_a(\mbox{\boldmath $k$})
H_a(\mbox{\boldmath $k$})
\right.
\nonumber \\
&&+
{\cal D}^{1*}_{\mu\nu}(\Xi)
{\cal D}^{1}_{\mu\nu'}(\Xi)
H^*_b(\mbox{\boldmath $k$})
H_b(\mbox{\boldmath $k$})
\nonumber \\
&&+
{\cal D}^{1*}_{\mu\nu}(\Omega)
{\cal D}^{1}_{\mu\nu'}(\Xi)
H^*_a(\mbox{\boldmath $k$})
H_b(\mbox{\boldmath $k$})
\nonumber \\
&&+
\left.
{\cal D}^{1*}_{\mu\nu}(\Xi)
{\cal D}^{1}_{\mu\nu'}(\Omega)
H^*_b(\mbox{\boldmath $k$})
H_a(\mbox{\boldmath $k$})
\right]
Y^{*}_{1\nu} (\hat{k}) Y_{1\nu'} (\hat{k})
\end{eqnarray}
After the angular average, we  obtain 
\begin{eqnarray}
W^{(1)}(\mu, \mbox{\boldmath $k$}) 
& =& \frac{1}{3}\left[
H^*_a(\mbox{\boldmath $k$}) H_a(\mbox{\boldmath $k$})+
H^*_b(\mbox{\boldmath $k$}) H_b(\mbox{\boldmath $k$})\right]
\nonumber \\
&=&\frac{1}{3} k^2 (G'(k^2))^2 \left[
(1+\frac{\mbox{\boldmath $v$} \cdot \mbox{\boldmath $k$}}
{\omega_k})^2
+(1-\frac{\mbox{\boldmath $v$} \cdot \mbox{\boldmath $k$}}
{\omega_k})^2\right]
\label{eq:one-body}
\end{eqnarray}
which is independent of $\mu$ and the impact parameter 
$\mbox{\boldmath $b$}$, 
as we expect of a one-body observable averaged over independent 
groomings.

Next we look at the 2-body rate, that is the probability 
of finding a pion of momentum $\mbox{\boldmath $k$}$ and 
charge type $\mu$ 
and a pion of momentum $\mbox{\boldmath $q$}$ 
and charge type $\rho$ 
in the state $|h\rangle$. We have 
\begin{eqnarray}
W^{(2)}(\mu, \mbox{\boldmath $k$}; \rho, 
\mbox{\boldmath $q$})
&=&\frac{1}{(8\pi^2)^2}\int d\Omega d\Xi
\; h_{\mu}^* (\mbox{\boldmath $k$},\Omega,\Xi)
\;h_{\mu} (\mbox{\boldmath $k$},\Omega,\Xi)
\; h_{\rho}^* (\mbox{\boldmath $q$},\Omega,\Xi)
\;h_{\rho} (\mbox{\boldmath $q$},\Omega,\Xi)
\nonumber \\
&=&\frac{1}{36\pi^2}\int d\Omega d\Xi
\nonumber \\
& & \left[ 
{\cal D}^{1*}_{\mu\nu}(\Omega)
{\cal D}^{1}_{\mu\nu'}(\Omega)
{\cal D}^{1*}_{\rho\sigma}(\Omega)
{\cal D}^{1}_{\rho\sigma'}(\Omega)
H^*_a(\mbox{\boldmath $k$})
H_a(\mbox{\boldmath $k$})
H^*_a(\mbox{\boldmath $q$})
H_a(\mbox{\boldmath $q$})
\right.
\nonumber \\
&&+
{\cal D}^{1*}_{\mu\nu}(\Omega)
{\cal D}^{1}_{\mu\nu'}(\Omega)
{\cal D}^{1*}_{\rho\sigma}(\Omega)
{\cal D}^{1}_{\rho\sigma'}(\Xi)
H^*_a(\mbox{\boldmath $k$})
H_a(\mbox{\boldmath $k$})
H^*_a(\mbox{\boldmath $q$})
H_b(\mbox{\boldmath $q$})
\nonumber \\
&&+
{\cal D}^{1*}_{\mu\nu}(\Omega)
{\cal D}^{1}_{\mu\nu'}(\Omega)
{\cal D}^{1*}_{\rho\sigma}(\Xi)
{\cal D}^{1}_{\rho\sigma'}(\Omega)
H^*_a(\mbox{\boldmath $k$})
H_a(\mbox{\boldmath $k$})
H^*_b(\mbox{\boldmath $q$})
H_a(\mbox{\boldmath $q$})
\nonumber \\
&&+
{\cal D}^{1*}_{\mu\nu}(\Omega)
{\cal D}^{1}_{\mu\nu'}(\Omega)
{\cal D}^{1*}_{\rho\sigma}(\Xi)
{\cal D}^{1}_{\rho\sigma'}(\Xi)
H^*_a(\mbox{\boldmath $k$})
H_a(\mbox{\boldmath $k$})
H^*_b(\mbox{\boldmath $q$})
H_b(\mbox{\boldmath $q$})
\nonumber \\
&&+
{\cal D}^{1*}_{\mu\nu}(\Omega)
{\cal D}^{1}_{\mu\nu'}(\Xi)
{\cal D}^{1*}_{\rho\sigma}(\Omega)
{\cal D}^{1}_{\rho\sigma'}(\Omega)
H^*_a(\mbox{\boldmath $k$})
H_b(\mbox{\boldmath $k$})
H^*_a(\mbox{\boldmath $q$})
H_a(\mbox{\boldmath $q$})
\nonumber \\
&&+
{\cal D}^{1*}_{\mu\nu}(\Omega)
{\cal D}^{1}_{\mu\nu'}(\Xi)
{\cal D}^{1*}_{\rho\sigma}(\Omega)
{\cal D}^{1}_{\rho\sigma'}(\Xi)
H^*_a(\mbox{\boldmath $k$})
H_b(\mbox{\boldmath $k$})
H^*_a(\mbox{\boldmath $q$})
H_b(\mbox{\boldmath $q$})
\nonumber \\
&&+
{\cal D}^{1*}_{\mu\nu}(\Omega)
{\cal D}^{1}_{\mu\nu'}(\Xi)
{\cal D}^{1*}_{\rho\sigma}(\Xi)
{\cal D}^{1}_{\rho\sigma'}(\Omega)
H^*_a(\mbox{\boldmath $k$})
H_b(\mbox{\boldmath $k$})
H^*_b(\mbox{\boldmath $q$})
H_a(\mbox{\boldmath $q$})
\nonumber \\
&&+
{\cal D}^{1*}_{\mu\nu}(\Omega)
{\cal D}^{1}_{\mu\nu'}(\Xi)
{\cal D}^{1*}_{\rho\sigma}(\Xi)
{\cal D}^{1}_{\rho\sigma'}(\Xi)
H^*_a(\mbox{\boldmath $k$})
H_b(\mbox{\boldmath $k$})
H^*_b(\mbox{\boldmath $q$})
H_b(\mbox{\boldmath $q$})
\nonumber \\
&&\left. +(a\leftrightarrow b, \Omega\leftrightarrow 
\Xi)\right] Y^{*}_{1\nu} (\hat{k}) Y_{1\nu'} (\hat{k})
Y^{*}_{1\sigma} (\hat{q}) Y_{1\sigma'} (\hat{q})
\end{eqnarray}
Terms with odd numbers of $\Omega$ or $\Xi$  
average to zero. 
Applying the following formulae for Wigner functions
\begin{equation}
\int   
{\cal D}^{1*}_{\mu\nu} 
{\cal D}^{1}_{\rho\sigma}= \frac{8\pi^2}{3}\delta_{\mu\rho}
\delta_{\nu\sigma},
\end{equation}
\begin{eqnarray}
\int 
{\cal D}^{1}_{\mu\nu} 
{\cal D}^{1}_{\rho\sigma}&=&
(-1)^{\mu-\nu} 
\int {\cal D}^{1*}_{-\mu-\nu}
{\cal D}^{1}_{\rho\sigma}
\nonumber \\
&=&
\frac{8\pi^2}{3}
(-1)^{\mu-\nu} 
\delta_{\mu,-\rho}
\delta_{\nu,-\sigma},
\end{eqnarray}
\begin{equation}
\int 
{\cal D}^{1*}_{\mu\nu} 
{\cal D}^{1*}_{\rho\sigma}
{\cal D}^{1}_{\mu\nu'} 
{\cal D}^{1}_{\rho\sigma'} 
= 8\pi^2\sum_{\Lambda M M'}
\frac{1}{2\Lambda+1} 
\langle 11;\mu\rho| \Lambda M \rangle ^2
\langle 11;\nu\sigma| \Lambda M' \rangle
\langle 11;\nu'\sigma'| \Lambda M' \rangle,
\end{equation}
the two-body rate is then
\begin{eqnarray} 
W^{(2)}(\mu, \mbox{\boldmath $k$}; \rho, \mbox{\boldmath $q$})
= \frac{1}{9} \left[ \right. &A_{\mu\rho}(z)&
(H_a^*(\mbox{\boldmath $k$}) 
H_a(\mbox{\boldmath $k$}) 
H_a^*(\mbox{\boldmath $q$}) 
H_a(\mbox{\boldmath $q$}) 
+(a\leftrightarrow b))
\nonumber \\
+&B_{\mu\rho}(z) &
(
H_a^*(\mbox{\boldmath $k$})
H_a(\mbox{\boldmath $k$})
H_b^*(\mbox{\boldmath $q$})
H_b(\mbox{\boldmath $q$})
+(a\leftrightarrow b))
\nonumber \\
+
&C_{\mu\rho}(z)&
(H_a^*(\mbox{\boldmath $k$})
H_b(\mbox{\boldmath $k$})
H_a^*(\mbox{\boldmath $q$})
H_b(\mbox{\boldmath $q$})
+(a\leftrightarrow b))
\nonumber \\
+
&D_{\mu\rho}(z)&
\left.
(H_a^*(\mbox{\boldmath $k$})
H_b(\mbox{\boldmath $k$})
H_b^*(\mbox{\boldmath $q$})
H_a(\mbox{\boldmath $q$})
+(a\leftrightarrow b))\right]
\label{eq:two-body}
\end{eqnarray}
where 
\begin{eqnarray}
A_{\mu\rho}(z) &=& 9\sum_{L\Lambda} (-1)^{L}
\langle 11;\mu\rho|LM\rangle^2\;
\langle 11;00|\Lambda 0\rangle^2\;
W(1111;L\Lambda) P_{\Lambda}(z) 
\label{eq:A}
\\
B_{\mu\rho}(z)& = & 1
\\
C_{\mu\rho}(z)&=& \delta_{\mu,-\rho}\; z^2
\\
D_{\mu\rho}(z)&=& \delta_{\mu,\rho} \; z^2
\end{eqnarray}
with $z=\hat{k}\cdot\hat{q}$.


Putting the explicit forms of $H_a$ and $H_b$ (\ref{eq:Hab})
into $W^{(2)}$, we have
\begin{eqnarray}
W^{(2)}
=&\frac{1}{9}&
k^2 (G'(k^2))^2
q^2 (G'(q^2))^2
\left[
\right.
\nonumber \\
&& A_{\mu\rho}(z)\, ((1+\frac{\mbox{\boldmath $v$} \cdot 
\mbox{\boldmath $k$}}{\omega_k})^2
(1+\frac{\mbox{\boldmath $v$} \cdot \mbox{\boldmath $q$}}{\omega_q})^2
+ (1-\frac{\mbox{\boldmath $v$} \cdot \mbox{\boldmath $k$}}{\omega_k})^2
(1-\frac{\mbox{\boldmath $v$} \cdot \mbox{\boldmath $q$}}{\omega_q})^2)
\nonumber \\
&+&  ((1+\frac{\mbox{\boldmath $v$} \cdot \mbox{\boldmath $k$}}{\omega_k})^2
(1-\frac{\mbox{\boldmath $v$} \cdot \mbox{\boldmath $q$}}{\omega_q})^2 +
(1+\frac{\mbox{\boldmath $v$} \cdot \mbox{\boldmath $q$}}{\omega_q})^2
(1-\frac{\mbox{\boldmath $v$} \cdot \mbox{\boldmath $k$}}{\omega_k})^2)
\nonumber \\
&+& \delta_{\mu,-\rho}z^2\, (1-(\frac{\mbox{\boldmath $v$} \cdot 
\mbox{\boldmath $k$}}{\omega_k})^2)
(1-(\frac{\mbox{\boldmath $v$} 
\cdot \mbox{\boldmath $q$}}{\omega_q})^2) 2 
\cos((\mbox{\boldmath $k$}+\mbox{\boldmath $q$})
\cdot \mbox{\boldmath $b$})
\nonumber \\
&+& 
\left.
\delta_{\mu\rho} z^2\, (1-(\frac{\mbox{\boldmath $v$} \cdot 
\mbox{\boldmath $k$}}{\omega_k})^2)
(1-(\frac{\mbox{\boldmath $v$}\cdot 
\mbox{\boldmath $q$}}{\omega_q})^2) 2 
\cos((\mbox{\boldmath $k$}-\mbox{\boldmath $q$})
\cdot \mbox{\boldmath $b$})\right] \label{eq:two-body-f}
\end{eqnarray}

The quantity $A_{\mu\rho}(z)$ (\ref{eq:A}) is easily evaluated to be
\begin{equation}
A_{\mu\rho}(z)=\left\{ 
\begin{array}{ll}
\frac{9}{10}+\frac{3}{10}z^2 & \mbox{ if $\mu\neq 0$ and $\rho\neq 0$} \\
\frac{6}{5}-\frac{3}{5}z^2  & \mbox{ if only $\mu$ or $\rho$ is zero} \\
\frac{3}{5}+\frac{6}{5}z^2 & \mbox{if $\mu= 0$ and $\rho=0$}
\end{array}
\right. .
\end{equation}

We should also observe that
\begin{equation}
A_{\mu\rho}(z)=A_{\rho\mu}(z), \;\;\; \sum_{\mu\rho} A_{\mu\rho}(z)= 9.
\end{equation}

There are two kinds of pion correlations studied in the literature.
The first is the standard correlation function  defined by
\begin{equation}
C_2(\mu, \mbox{\boldmath $p$};\rho, 
\mbox{\boldmath $q$}) = \frac{W_2(\mu, 
\mbox{\boldmath $p$};\rho, \mbox{\boldmath $q$})}
{W_1(\mu,\mbox{\boldmath $p$}) W_1(\rho, \mbox{\boldmath $q$})}-1
\label{eq:C2}
\end{equation}
and the second form, most used in extracting pion 
correlations from annihilation data, is
\begin{equation}
R_2(\mu, \mbox{\boldmath $p$};\nu, 
\mbox{\boldmath $q$}/(\rho, \mbox{\boldmath $p$};\sigma, 
\mbox{\boldmath $q$} ) = 
\frac{W_2(\mu, \mbox{\boldmath $p$};\nu, \mbox{\boldmath $q$})}
{W_2(\rho, \mbox{\boldmath $p$};\sigma, \mbox{\boldmath $q$})}.
\label{eq:R2}
\end{equation}
This form is suitable for studying correlations from 
many random sources, where the two-body rates for different 
type of particles (like $\pi^+\pi^-$) approaches the product of 
the corresponding one-body rates. This form is commonly used
for ratios of correlations among like to unlike
pions (eg $\mu = \nu$).

Annihilation occurs over a range of the impact parameter 
$\mbox{\boldmath $b$}$.
Substantial annihilation is expected for $ b\le 1$fm while 
it should be minimal for $b$ much large than $1$fm.
To further examine the correlations, we average 
over the impact parameter $\mbox{\boldmath $b$}$ with the probability 
$e^{-\lambda b }$. Here $\lambda$ is a range parameter of the
order of one inverse fermi, specifying
the significant region of collision of Skyrmion 
and anti-Skyrmion. In the 
two-body rate (\ref{eq:two-body-f}) only the last two terms depend
on $\mbox{\boldmath $b$}$. The integral needed for the average is 
\begin{equation}
\int d^2b\, e^{-\lambda b} \cos(\mbox{\boldmath $Q$}\cdot 
\mbox{\boldmath $b$}) = 
\int db\,d\phi\, b\, e^{-\lambda b} \cos(Q_{\bot} b \cos\phi )
\label{eq:bave}
\end{equation}
where $\mbox{\boldmath $Q$}=\mbox{\boldmath $k$}\pm
\mbox{\boldmath $q$}$, $Q_{\bot}$ is the magnitude of
the  component of $\mbox{\boldmath $Q$}$ in the 
$\mbox{\boldmath $b$}$ plane, and 
$\phi$ is the angle between this component and $\mbox{\boldmath $b$}$. This
integral gives 
\begin{equation}
\frac{2\pi\lambda}{(\lambda^2+ Q_{\bot}^2)^{\frac{3}{2}}}.
\end{equation}
The result of such an averaging is achieved simply by
replacing $\cos(\mbox{\boldmath $Q$}\cdot\mbox{\boldmath $b$})$ 
by $\lambda^3/(\lambda^2+
Q_{\bot}^2)^{\frac{3}{2}}$ in (\ref{eq:two-body-f}).

\section{Lorentz transformed hedgehog}

When the velocity of the moving hedgehog is not small compared to
the speed of light, the Lorentz transformed field configuration should be 
used.  In the overall CM frame, the moving hedgehog (in the direction
of positive x-axis)
$\vec{\phi}_a$ in (\ref{eq:SantiS}) can be written as
\begin{equation}
\vec{\phi}_a= (\gamma(x-vt), y+b/2,z) 
\gamma^2g((x-vt)^2+
(y+b/2)^2+z^2) \label{eq:Srela}
\end{equation}
considering that the pion field transforms as a pseudo-scalar. Here 
$\gamma=1/\sqrt{1-v^2/c^2}$. Similarly
the Lorentz transformed anti-Skyrmion is 
\begin{equation}
\vec{\phi}_b= -(\gamma(x+vt), y-b/2,z) g(\gamma^2 (x+vt)^2+
(y-b/2)^2+z^2) \label{eq:antiSrela}
\end{equation}

Compare (\ref{eq:Srela}), (\ref{eq:antiSrela}) with the Galilean boosted
field configurations \ref{eq:SantiS} and their Fourier transforms
\ref{eq:Fourier1}, we define
the following ``capitalized" quantities
\begin{eqnarray}
\mbox{\boldmath $V$}&=&\gamma \mbox{\boldmath $v$}\nonumber \\
\mbox{\boldmath $R$}&=&(\gamma x,y, z) \nonumber \\
\mbox{\boldmath $K$}&=&(k_x/\gamma,k_y, k_z) .
\end{eqnarray}
The calculation of the fourier amplitudes for the Lorentz case is 
essentially the same as in the Galilean case, if we 
replace the lower case vectors  in the Galilean case 
by these ``capitalized" vectors.
We should  also note that $d^3 r=\frac{1}{\gamma} d^3R$.

The Fourier amplitudes for the Skyrmion and antiSkyrmion are 
then
\begin{eqnarray}
\frac{1}{\gamma} \vec{h}_a (\mbox{\boldmath $K$}, \omega_k) &= & 
i\frac{\mbox{\boldmath $K$}}{\gamma}
e^{i\mbox{\boldmath $k$}\cdot{\frac{\mbox{\boldmath $b$}}{2}}}G'(K^2)
(1+\frac{\bf{v}\cdot\bf{k}}{\omega_k})
\\
\frac{1}{\gamma} \vec{h}_a (\mbox{\boldmath $K$}, \omega_k) &= & 
-i\frac{\mbox{\boldmath $K$}}{\gamma}
e^{i\mbox{\boldmath $k$}\cdot{\frac{\mbox{\boldmath $b$}}{2}}}G'(K^2)
(1-\frac{\bf{v}\cdot\bf{k}}{\omega_k})
\end{eqnarray}

Independent random grooming on the (anti)hedgehog is 
simply a rotation on the vector $\mbox{\boldmath $K$}$ for the Lorentz
case. Consequently, the one-body and two body rates,
(\ref{eq:one-body}) and 
(\ref{eq:two-body}), are replaced by
\begin{equation}
W^{(1)}_{L}(\mu,\mbox{\boldmath $k$})=
\frac{1}{\gamma^2}
W^{(1)}_{G}(\mu,\mbox{\boldmath $K$})
\end{equation}
and
\begin{equation}
W^{(2)}_{L}(\mu,\mbox{\boldmath $k$}; \nu, \mbox{\boldmath $q$})
=
\frac{1}{\gamma^4}
W^{(2)}_{G}(\mu,\mbox{\boldmath $K$}; \nu, \mbox{\boldmath $Q$})
\end{equation}
Here $L$ stands for Lorentz and $G$ Galilean. The
cosine of the relative  angle in (\ref{eq:two-body}) should
be replaced by $\hat{K}\cdot\hat{Q}$. In the plane perpendicular
to the velocity, the two cases give identical pion correlations.

\section{Result and Discussion}

Pion correlations arise from the 
incoherence in emission sources, commonly
from thermal or experimental averages. 
In the present case, however, the Skyrme 
dynamics (in particular, the coupling
between spatial and isospin degrees of 
freedom)  also leads to interesting correlations.
Although related to the regular 
type of correlations\cite{ACDLL}, the present
study points to correlations not only among 
pions of like charges but among unlike charges
as well. Our present consideration does
not include full dynamical evolution 
nor final state interactions among the pions.
However, we hope important aspects of the Skyrme 
physics in $N\overline{N}$ annihilation 
are elucidated here and the qualitative features will
remain in a more thorough study. 

As an example of what our picture gives for correlations, we
study the $R_2$ of (\ref{eq:R2}).  This is the correlation function
most commonly studied in annihilation.  What is usually 
studied is not the full $R_2$ as a function of both
momentum variables, but rather $R_2$ as a function of
pion pair relative momentum summed over all total 
momenta.  That is we define 
$\mbox{\boldmath $Q$} = \mbox{\boldmath $p$} - 
\mbox{\boldmath $q$}$
and $\mbox{\boldmath $P$} = \mbox{\boldmath $p$} 
+ \mbox{\boldmath $q$}$.  Then consider $R_2$ for
fixed $Q$ summed over all $\mbox{\boldmath $P$}$ as well 
as being averaged over the random groomings
and impact parameter.  Note that after those averages, $R_2$ is
a function of the magnitude of $Q$ only.  We study $R_2$ using the
Lorentz boosted form for two values of the incident energy, or
velocity, $v=0.1 c$ and $v=0.9 c$. In the sum over 
$\mbox{\boldmath $P$}$  
for fixed $\mbox{\boldmath $Q$}$ we take into account the constraints of
energy and momentum conservation.  In order to do this calculation
we need to choose a form for $G(k^2)$ defined in (15).  We 
take the simple form, $G(k^2) = 1/(k^2+\mu^2)$, 
and take $\mu = \lambda = 1$fm$^{-1}$, where $\lambda$ is the range
in the impact parameter average.
We study $R_2$ as a function of $Q$ for various charge ratios.
These are shown in Figure 2.  For all cases $R_2$ tends to
one for large $Q$, reflecting the absence of charge correlations
for large relative momentum.  The range over which $R_2$ differs
significantly from one is $1$fm$^{-1}$ and is set by our choice
of scale.  In all cases the like charge pion to unlike charge
correlations resemble closely the correlations seen in the
data and usually attributed to Bose-Einstein effects. Note
that we see them here although we have made no
statistical or thermal assumptions.
As we have discussed before \cite{ACDLL}, we have averaged
over $\mbox{\boldmath $P$}$, impact parameter and grooming.

Similar correlations (especially among unlike charged pions) has
been discussed \cite{BD} for heavy ion collisions as a consequence of
Skyrme dynamics. However, it may be easier to examine these correlations
experimentally in  $N\overline{N}$ annihilation since 
the dynamics is much simpler. In further extending
our study of the manifestation of large QCD in annihilation, we 
need to study quantum corrections, time 
dependent rotational Skyrmions and their collisions, 
interactions among the emitted pion waves and the construction
of a coherent state representing a half integer spin soliton.
These considerations may hold the clue to 
two-pion charge-spin asymmetry \cite{Hasan} and 
other interesting experimental findings \cite{Myhrer}. 
These studies will extend the domain of the classical/quantum 
approach to annihilation physics. \\

This work is partially supported by the United
States National Science Foundation.

\clearpage

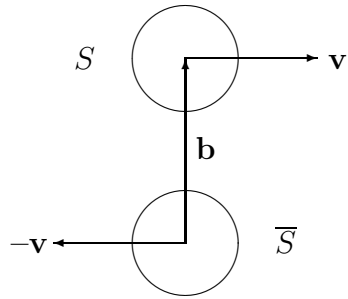
\begin{figure}
\begin{picture}(150,200)(10,20)
\put(210,160){\circle{40}}
\put(210,90){\circle{40}}
\put(168,156){$S$}
\put(244,86){$\overline{S}$}
\put(210,90){\vector(0,1){70}}
\put(210,90){\vector(-1,0){50}}
\put(210,160){\vector(1,0){50}}
\put(214,122){{\bf b}}
\put(264,156){{\bf v}}
\put(144,86){{\bf --v}}
\end{picture}
\caption{Skyrmion and antiSkyrmion in collision.}
\end{figure}

\newpage

\begin{figure}
\centerline{\hbox{
\psfig{figure=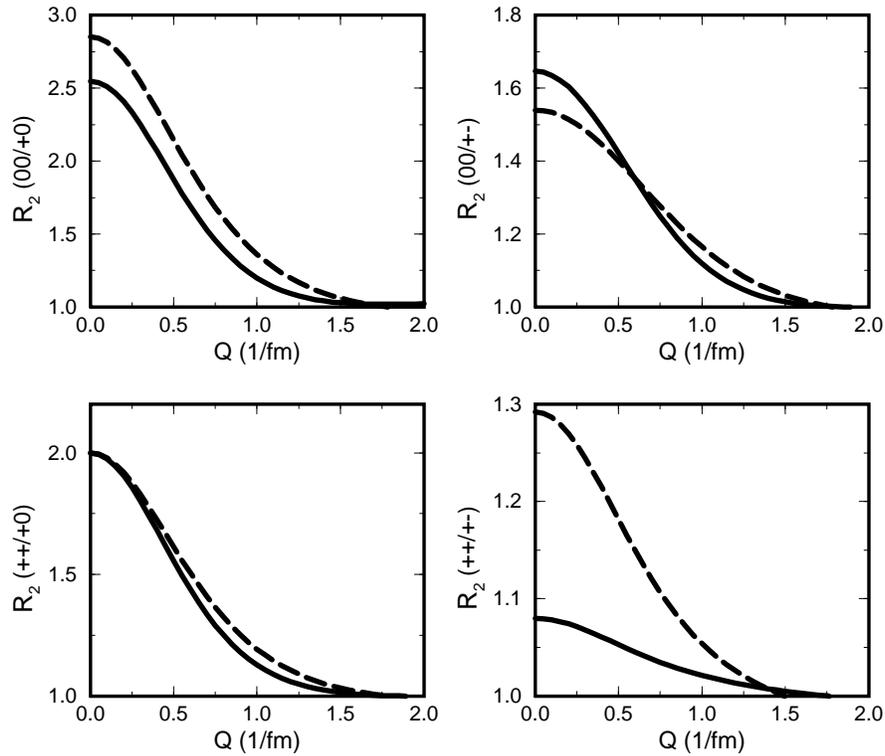,height=4.7in} }}
\caption{The two pion correlation, $R_2$ as defined 
in (\protect{\ref{eq:R2}}),  as a function of pion relative momentum 
$Q$ after grooming and averaging over total momentum  of the 
two pions and the direction of antiproton velocity 
(as discussed in the main text). We show here $R_2$ for different charge 
combinations. Solid curves are for $v=0.1c$ and dashed for $v=0.9c$.
}
\end{figure}

\end{document}